\documentstyle[12pt]{article}
\begin{document}
\begin{titlepage}

\begin{flushright}
ISAS/EP/95/101\\
August 1995
\end{flushright}
\vspace*{0.5cm}

\begin{center}
{\bf
\begin{Large}
{\bf
Random Walk with a Boundary Line as a \\
Free Massive Boson with
a Defect Line
\\}
\end{Large}
}
\vspace*{1.5cm}
 {\large A. Valleriani} \footnote{E-mail: angelo@frodo.sissa.it}
         \\[.3cm]
          {\it International School for Advanced Studies}\\
          {\it and}\\
          {\it Istituto Nazionale di Fisica Nucleare}\\
          {\it 34013 Trieste, Italy}\\
\end{center}
\vspace*{0.7cm}

\begin{abstract}
\noindent
We show that the problem of Random Walk with boundary attractive
potential may be mapped onto the free massive bosonic Quantum Field
Theory with a line of defect. This mapping permits to recover the
statistical properties of the Random Walks by using boundary $S$--matrix
and Form Factor techniques.
\end{abstract}
\vfill
\end{titlepage}
\setcounter{footnote}{0}
\section{Introduction}

The problem of the Random Walk in presence of a boundary line near the
so--called {\it compensation point} for long chains has been solved years
ago by using standard statistical mechanics methods (see \cite{EKB} and
references therein). It is nevertheless worth to reconsider this model in
the light of recent developments in boundary Quantum Field Theory
\cite{goza}, in order to understand in a deeper way the connection between
the classical configurations of chains and Green's functions in the
corresponding Quantum Field Theory model. In this paper, we will show that
the statistical problem of the 2-d Random Walk with a boundary line can be
mapped onto a bosonic Quantum Field Theory with a defect line. Namely, we
will see that in order to derive the statistical behaviour of the Random
Walk in the presence of a boundary condition, one has to treat the
boundary not as a pure classical object but as a quantum defect line in
the corresponding free massive boson model, where both Reflection and
Transmission amplitudes are needed. As a by-product of our results, we
show that the sum of the aforementioned amplitudes plays
the role of the
boundary $S$--matrix for the
free massive bosonic Quantum Field Theory in
half--plane, such that a definition of a boundary state for this problem
can be used to compute the quantities we are interested in.

The Quantum Field Theory approach presented in this paper may be useful to
analyse the analogous problem with the Random Walk substituted by the Self
Avoiding Walk. We would like to remind that in the bulk, many
geometrical quantities of the Self Avoiding Walk can be obtained by using
an $S$--matrix approach \cite{Zam1,cm}, relying on the relationship
between Self Avoiding Walks and the $O(n)$ model for $n\rightarrow 0$
\cite{deg}. This relationship has already been used to discuss several
interesting aspects in the presence of a boundary condition\footnote{
Fendley and Saleur \cite{FS} have recently conjectured the exact boundary
$S$--matrix for the Self Avoiding Walk, by using an analogy with the
corresponding amplitude of the {\it Kondo problem}. It would be
interesting to have a direct derivation of this quantity as a solution of
the functional equations satisfied by the boundary $S$--matrix.}.

\section{ The Random Walk with boundary}

In this section, we will review some results of the Random Walk with
boundary in order to establish the correspondence with the language of
Quantum Field Theory. We closely follow the formulation given in
ref.~\cite{Eis} (for the problem in the bulk, see \cite{ItzD} and
\cite{MonWest}).

Let us initially consider the simplest model: the one--dimensional Random Walk
on the lattice, with the walker confined to move only on the positive
half--line $x\geq 0$. With a potential
\begin{equation}
V=\left\{ \begin{array}{ll}
\epsilon & \mbox{if $x=0$} \\
0 & \mbox{if $x\geq 1$},
\end{array}
\right.
\end{equation}
the partition function
for the configurations is given by
\begin{equation}
Z_V(x,x_0;N)=\sum_{n_0=0}^{\infty}\,a^{n_0}\, Z(x,x_0;N;n_0),
\label{parv}
\end{equation}
where $a\sim e^{-\epsilon/kT}$ and $n_0$ is the number of times a given path
{\it sits} in the origin. The partition function $Z$
in the rhs of (\ref{parv})
counts the number of different configurations, in the absence of potential,
of a chain of length $N$ with fixed ends ($x$, $x_0$) and which sits
$n_0$ times in the origin. By using the {\it images method} \cite{Cha}, this
expression can be reduced to
\begin{equation}
Z_V(x,x_0;N)= Z_b(x,x_0;N;n_0=0) +
2\sum_{n_0=1}^{\infty}\,\left(\frac{a}{2}\right)^{n_0}\,Z(x,x_0;N;n_0) \, ,
\label{para}
\end{equation}
where $Z_b$ is the partition function in the bulk.

For $\epsilon<0$, there exists a critical temperature $T_c$ such that for
$T=T_c$ we get $a_c=2$. This value of the temperature defines the
so--called {\it compensation point}, where the walker does not feel any
driving force, neither the (entropic) repulsion nor the (energetic)
attraction. In fact, for $T>T_c$, we observe a preference for the walker
to escape from the potential well, i.e. the favourite configurations are
those which end far away from the boundary. This will be called the {\it
non-adsorbed phase} of the Random Walk. On the contrary, for $T<T_c$ the
favourite configurations are those approaching the boundary with a low
probability to escape. This will be called the {\it adsorbed phase}.

The existence of two distinct phases of the Random Walk and a critical
point in between can be also established in the case of two--dimensional
Random Walk with boundary \cite{Eis}. In the continuum limit,
in order to mimic the boundary around
the {\it compensation point}, the potential can be chosen as
\begin{equation}
W=\left\{ \begin{array}{ll}
\infty & \mbox{if $x\leq 0$} \\
< 0 & \mbox{if $0< x < b$} \\
0 & \mbox{if $x\geq b$}
\end{array}
\right.
\end{equation}
and independent from the coordinate
parallel to the boundary line, say $y$. Since the two--dimensional
partition function of the Random
Walk can be factorized into the product of two independent one--dimensional
partition functions, we will study first the one--dimensional problem and
then we will come back to the original two--dimensional case.

The Green's function of the Random Walk, given by the Laplace transform of
the partition function $Z$, in the one--dimensional case is the solution
of the differential equation
\begin{equation}
\label{defG}
\left(-\partial^2_x + m^2 +W(x)\right)G(x,x_0;m^2)=\delta(x-x_0),
\end{equation}
with the additional condition that it vanishes at infinity.
The above differential equation can be solved by using standard methods
(see for example \cite{smir}). Here we concentrate our attention on the
solution given by
\begin{equation}
\label{grf}
G(x,x_0;m^2)=\frac{e^{-m|x-x_0|}+F(m,b,T)e^{-m(x+x_0)}}{2m}
\end{equation}
for $x, x_0\geq0$ where all informations about the potential are encoded
into the function $F$. This function can be cast into the following
universal form \cite{EKB} (see also \cite{deg1})
\begin{equation}
\label{F} F=\frac{1-c/m}{1+c/m}\, ,
\end{equation} provided that the
length of the chain $\sqrt{N}>>b$ and that the function $c\propto (T-T_c)$
satisfies\footnote{The lower bound for $c$ negative comes from the
requirement that the denominator of (\ref{F}) be bigger than zero.  We are
supposing to have chosen a potential $W$ that satisfies these properties
\cite{Eis}.} $-b^{-1}<<c\leq b^{-1}$. Since we are interested in the
universal behaviour of the Random Walk chains, we may let at this point
$b\rightarrow 0$ and consider the Green's function (\ref{grf}) with the
above function $F$ as meaningful expressions for any $x,x_0\geq 0$. This
obviously implies that we are not concerned, from now on, with a microscopic
analysis of the interaction, much like in the spirit of the $S$--matrix
approach for the particle models.

We note here the following limits:
\begin{description}
\item[a)]
\noindent
for $c\rightarrow +\infty$, the function $F\rightarrow -1$
and the Green's function becomes
\[
G(x,x_0;m^2)=\frac{e^{-m|x-x_0|}-e^{-m(x+x_0)}}{2m} \,\,\, .
\]
This limit corresponds to the {\it hard--wall} behaviour for
$x,x_0$ far away from the boundary.

\item[b)]
\noindent
for $c\rightarrow 0$, we have instead
\[G(x,x_0;m^2)=\frac{e^{-m|x-x_0|}+e^{-m(x+x_0)}}{2m}\, .\]
This allows us the identification of the point $c=0$ in this description
as the {\it compensation point} of the Random Walk with boundary.
\end{description}

Finally, it is important to notice that the Green's function (\ref{defG}) of
the one--dimensional Random Walk can be also obtained as solution of the
{\it free} differential equation in one--dimension \cite{Eis}
\begin{equation}
\label{defGf}
\left(-\partial^2_x + m^2 \right)G(x,x_0;m^2)=\delta(x-x_0)\, ,
\end{equation}
but with the interaction encoded into the boundary condition
\begin{equation}
\label{boc}
\partial_x\,G(0,x_0;m^2)=c\,G(0,x_0;m^2)\, .
\end{equation}

Once the solution of the one--dimensional case has been obtained, the
Green's function of the two--dimensional Random Walk can be computed by
using Fourier transform as
\begin{equation}
G({\bf r},{\bf r}_0;m^2)=\int_{-\infty}^{+\infty}\frac{\mbox{d}k}{2\pi}\,
e^{ik(y-y_0)}\,G(x,x_0;m^2+k^2),
\end{equation}
where $G(x,x_0;m^2+k^2)$ satisfies the differential equation (\ref{defG}) and
is given by (\ref{grf}) and (\ref{F}) with the substitution
$m^2\rightarrow m^2+k^2$. This integral can be cast in the
suitable form
\begin{equation}
\label{gtot}
\begin{array}{c}
\begin{displaystyle}
G({\bf r},{\bf r}_0;m^2)=\int_0^{\infty}\frac{\mbox{d}\theta}{2\pi}\,
e^{im(y-y_0)\sinh\theta}\left(e^{-m|x-x_0|\cosh\theta} +\right.
\end{displaystyle}
\\
\\
\begin{displaystyle}
\left. +\hat{S}(\theta,m,c)\, e^{-m(x+x_0)\cosh\theta}\right)\, ,
\end{displaystyle}
\end{array}
\end{equation}
where
\begin{equation}
\label{bs}
\hat{S}(\theta,m.c) = \frac{\cosh\theta -c/m}{\cosh\theta +c/m}.
\end{equation}
{}From the equations (\ref{defGf}) and (\ref{boc}) satisfied by the
one--dimensional Green's function, it is simple to derive the differential
equation satisfied by the above one
\begin{equation}
\left(-\Delta_{\mbox{\bf r}} + m^2 \right)G(\mbox{\bf r},\mbox{\bf r}_0;m^2)
=\delta(\mbox{\bf r}-\mbox{\bf r}_0)
\end{equation}
supplied with the boundary condition
\begin{equation}
\left. \partial_x G({\bf r},{\bf r}_0;m^2)\right|_{x=0}=
\left. c\,G({\bf r},{\bf r}_0;m^2)\right|_{x=0}.
\end{equation}
Notice that the above equations are those satisfied by the two-point
correlation function for the euclidean massive boson with action
\begin{equation}
\label{action}
S[\varphi]=\int\,{\mbox d}x\,{\mbox{d}}y\,
\left\{\theta(x)\,
\left( \frac{1}{2}(\nabla\varphi)^2+\frac{m^2}{2}\varphi^2\right) +
\frac{c}{2}\delta(x)\varphi^2\right\},
\end{equation}
where $\theta(x)$ is the Heaviside distribution.

\section{The Quantum Field Theory approach}

Aim of this section is to show that there exists a one-to-one correspondence
between the problem of two--dimensional Random Walk with boundary near the
compensation point
and a Quantum Field Theory of a bosonic field $\varphi$
with a line of defect.
In particular, we will show that the pure {\it hard--wall}
situation in the Random Walk ($T\rightarrow\infty$) is described in terms
of a {\it totally reflective} defect in the Quantum Field Theory
model, while the {\it compensation point} ($T=T_c$) of the Random Walk
corresponds to a
{\it totally transmitting} defect, provided that the (classically) forbidden
negative half line is mirrored to the positive axis.

In order to establish this correspondence, the first step is to associate
to each chain of the Random Walk problem a trajectory of the particle
field $\varphi$ described by the Quantum Field Theory\footnote{In the context
of polymer
physics, this interpretation has been proposed in \cite{Zam1}.}. The second
step consists in solving a combinatorial problem arising from the counting of
the configurations. To this aim,
it will be convenient to consider two copies of the Random Walk problem,
defined on the left and right sides of the boundary respectively. The two
copies are subjected to the same potential well and share the same
temperature. In this picture, the boundary may be treated as a defect line.
Notice that, since at the {\it compensation point} the behaviour of the
Random Walk is like that in the absence of boundary, this corresponds, in the
two-copy scheme, to trajectories that start e.g. from the right side of the
boundary and end to the left side of it or viceversa. Said in other words,
the {\it compensation point} is mapped into the {\it pure transmitting}
behaviour of the defect line. Viceversa, the {\it hard--wall} limit
of the Random Walk corresponds to {\it purely reflecting} scattering
processes at the defect line.

Let us formulate more precisely this mapping. Consider the following action
\begin{equation}
S[\varphi_L,\varphi_R]=S[\varphi_L]+S[\varphi_R]
\label{action1}
\end{equation}
where
\[
S[\varphi_R]=\int\,{\mbox d}x\,{\mbox d}y\,\left\{\theta(x)\,\left(
\frac{1}{2}(\nabla\varphi_R)^2+\frac{m^2}{2}\varphi_R^2\right) +
\frac{c}{2}\delta(x)\varphi_R^2\right\}\]
and
\[S[\varphi_L]=\int\,{\mbox d}x\,{\mbox d}y\,\left\{\theta(-x)\,\left(
\frac{1}{2}(\nabla\varphi_L)^2+\frac{m^2}{2}\varphi_L^2\right) +
\frac{c}{2}\delta(x)\varphi_L^2\right\}.\]
The fields $\varphi_{L,R}$ are not independent since are linked each other
by the equation
\begin{equation}
\varphi_R(y,x)=\varphi_L(y,-x)\, .
\label{symc}
\end{equation}
The equations of motion associated to the action (\ref{action1}) are given by
\begin{equation}
\begin{array}{c}
\theta(x)\left(-\nabla_x + m^2\right)\varphi_R=0\\
\\
\theta(-x)\left(-\nabla_x + m^2\right)\varphi_L=0\\
\\
\left. \partial_x\left(\varphi_R-\varphi_L\right)\right|_{x=0}=
\left. c\,\left(\varphi_R+\varphi_L\right)\right|_{x=0}\\
\\
\varphi_R(y,0)=\varphi_L(y,0)
\end{array}
\end{equation}
where the last equality comes from equation (\ref{symc}).
Now we are in the position to see that this set of equations are the
euclidean version of those solved by A.J. Bray and M.A. Moore in \cite{BM},
who computed the Green's function (\ref{gtot}), and by G. Delfino,
G. Mussardo and P. Simonetti for the problem of the free relativistic
massive boson with a {\it line of defect} \cite{dms}. As proved in \cite{dms},
the dynamics of the massive boson with a line of defect is constrained by
the integrability conditions and is completely encoded into a set of
Transmission and Reflection amplitudes associated to the scattering processes
of the particle hitting the defect (Figure 1). Their explicit expressions are
given by \cite{dms}
\begin{equation}
\label{TR}
\begin{array}{c}
\displaystyle{
T(\beta,c)= \frac{\sinh\beta}{\sinh\beta+ic/m}}\\
\\
\displaystyle{
R(\beta,g)=-\frac{ic/m}{\sinh\beta+ic/m}}
\end{array}
\end{equation}
where now $\beta$ is the rapidity variable defined through the identity
\[ (E,p)=(m\cosh\beta,m\sinh\beta).\]

The remaining part of this letter will be devoted to the computation of the
two--point correlation function of the field $\varphi$ in the presence
of the defect line and to show that this quantity gives rise to the
Green's function (\ref{gtot}) of the Random Walk problem.

The correlation functions of the bosonic field $\varphi$ can be computed
by using the Form Factor approach for the integrable models \cite{smi,KW}.
This can be conveniently done by considering
the model defined in a geometry where the boundary or the defect are placed
at $t=0$. In this geometry, the boundary or defect line are promoted to
quantum operators which acts on the vacuum of the
bulk quantum theory whereas
the matrix elements of the fields remain those given in the bulk
case.

The Form Factors come from the insertion, in the correlation functions of a
given operator ${\cal O}({\bf r}),$ of a complete set of asymptotic states
\[
|\theta_1\cdots\theta_n>=A^{\dagger}(\theta_1)\cdots A^{\dagger}(\theta_n)|0>
\]
in such a way that for the time-ordered product of two such operators we have
\[
\begin{array}{l}
<0|{\cal O}({\bf r}_2){\cal O}({\bf r}_1)|0>=
\begin{displaystyle}
\sum_{n=0}^{\infty}\frac{1}{n!}\int\frac{{\mbox{d}}\theta_1
\cdots{\mbox{d}}\theta_n}{(2\pi)^n}|F_{{\cal{O}}}(\theta_1,\ldots,\theta_n)|^2
\cdot
\end{displaystyle}
\\
 \\
\cdot \exp\left[im(y_2-y_1)\sum_{i=0}^n
\sinh\theta_i-im|t_2-t_1|\sum_{i=0}^n\cosh\theta_i\right].
\end{array}
\]
The Form Factors are defined by
\[
F_{{\cal{O}}}(\theta_1,\ldots,\theta_n)=
<0|{\cal O}(0)|\theta_1\cdots\theta_n>\, .
\]
In our case, the creation and annihilation operators of the particle states
satisfy the usual bosonic commutation relations
\[
\left[A(\theta),A^{\dagger}(\beta)\right]=2\pi\delta(\theta-\beta)\, ,
\]
and this drastically simplifies the calculation of the Form Factors. In
fact, for the fields $\varphi_{L,R}$ we have
\begin{equation}
<0|\varphi(0)|\theta_1\cdots\theta_n>=\frac{1}{\sqrt 2}\delta_{n,1}
\end{equation}
while all the other non--vanishing Form Factors can be
computed by using Wick's theorem based on the algebra of the operators
$A(\theta)$ and $A^{\dagger}(\theta)$. With the above matrix elements, the
euclidean correlation function in the bulk is given by
\begin{equation}
<0|{\cal T}\left[\varphi(y,t)\varphi(y_0,t_0)\right]|0>_E=
\int_0^{\infty}\,\frac{\mbox{d}\theta}{2\pi}\,e^{-m|x-x_0|\cosh\theta
+im(y-y_0)\sinh\theta}\, ,
\end{equation}
where in the rhs we have set $it=x$.
Let us now consider the problem of computing correlation functions in the
presence of the defect line.
By considering the
scattering processes as occur at the defect line in their crossed channels
(Figure 2.a, 2.b), we need the new amplitudes given by
\begin{equation}
\begin{array}{l}
\hat{T}(\theta,c)=T(i\frac{\pi}{2}-\theta,c)=
\displaystyle{
\frac{\cosh\theta}
{\cosh\theta+c/m}} \\
\\
\hat{R}(\theta,c)=R(i\frac{\pi}{2}-\theta,c)=\displaystyle{
\frac{-c/m}{\cosh\theta+c/m}}\, .
\end{array}
\end{equation}
The computation of the correlation functions in presence of the defect
operator ${\cal{D}}$ can be performed by using the equations
\[
<\varphi(y_1,t_1)\cdots\varphi(y_n,t_n)>=
\frac{<0|{\cal{T}}[\varphi(y_1,t_1)\cdots{\cal{D}}\cdots\varphi(y_n,t_n)]|0>}
{<0|{\cal{D}}|0>}
\]
where the matrix elements of the operator ${\cal D}$ are given by
\cite{dms}
\[
\begin{array}{c}
<\beta|{\cal{D}}|\theta>=2\pi\hat{T}(\beta,c)\delta(\beta-\theta)\\
\\
<\beta_1,\beta_2|{\cal{D}}|0>=2\pi\hat{R}(\beta,c)\delta(\beta_1+\beta_2)
\end{array}
\]
and $<0|{\cal{D}}|0>=1.$

There are two cases to consider: the first case is when the two operators
$\varphi$ are across the defect line and the second one is when both fields
are located on the same side with respect the defect line.
With the Wick rotation $it=x$ and the defect line placed at $x=0$,
in the first case we have
\begin{equation}
\begin{array}{c}
<0|\varphi(y,x)\,{\cal{D}}\,\varphi(y_0,-x_0)|0>_E= \\
\\
\begin{displaystyle}
=\int_0^{\infty}\frac{\mbox{d}\theta}{2\pi}e^{-m(x+x_0)\cosh\theta}
e^{im(y-y_0)\sinh\theta}\hat{T}(\theta,c)\, ,
\end{displaystyle}
\end{array}
\end{equation}
whereas in the second case
\begin{equation}
\label{rco}
\begin{array}{c}
\begin{displaystyle}
<0|\varphi(y,x)\varphi(y_0,x_0)\,{\cal{D}}|0>_E=
\int_0^{\infty}\frac{\mbox{d}\theta}{2\pi}\left(e^{-m|x-x_0|\cosh\theta}
e^{im(y-y_0)\sinh\theta}+\right.
\end{displaystyle}
\\
\\
\begin{displaystyle}
\left. +e^{-m(x+x_0)\cosh\theta}
e^{im(y-y_0)\sinh\theta}\hat{R}(\theta,c)\right).
\end{displaystyle}
\end{array}
\end{equation}
The sum of the two contributions
\begin{equation}
\label{gfin}
\begin{array}{c}
G({\bf r},{\bf r}_0;m^2)=
<0|\varphi(y,x)\,{\cal{D}}\,\varphi(y_0,-x_0)|0>_E+ \\
\\
 + <0|\varphi(y,x)\varphi(y_0,x_0)\,{\cal{D}}|0>_E
\end{array}
\end{equation}
is exactly the required Green's function (\ref{gtot}).

Notice that there is another way to compute the same quantity: in fact,
one could mirror the left half plane {\it ab initio} to the right one and
consider the Transmission amplitude as if it was a sort of Reflection
amplitude. One can use this observation in order to define the function
\begin{equation}
K(\theta)=\hat{R}(\theta)+\hat{T}(\theta)
\end{equation}
which together with its crossed counter part defined as
\begin{equation}
\label{KR}
K(\theta)=R\left(\frac{i\pi}{2}-\theta\right)
\end{equation}
satisfy all the conditions (boundary Yang--Baxter, boundary unitarity and
boundary cross--unitarity) a boundary $S$--matrix should fulfill \cite{goza}.

It is thus possible to define the boundary state at the euclidean time
$x=0$
\begin{equation}
\label{bst}
|B>=\exp\left[\frac{1}{2}\int_{-\infty}^{+\infty}
\frac{\mbox{d}\theta}{2\pi}K(\theta)A^{\dagger}(-\theta)
A^{\dagger}(\theta)\right]|0>
\end{equation}
by which we may rewrite our Green's function as
\begin{equation}
\label{last}
G({\bf r},{\bf r}_0;m^2)=
<0|\varphi(y,x)\varphi(y_0,x_0)|B>_E\, .
\end{equation}

\section{Concluding remarks}

Although the original problem of the Random Walk near the compensation point
is defined in half-space, it has been convenient to formulate the dynamics in
terms
of two copies defined for $x>0$ and $x<0$ with appropriate
boundary conditions. In particular, it has been possible to identify the
Transmission amplitude of the defect line model with the compensative role of
the potential in the Random Walk problem and the Reflection amplitude of the
defect line with the {\it hard--wall} limit. For a generic temperature,
the dynamics is ruled by an overlapping of the two contributions, as shown
in equation (\ref{gfin}).

Note that the Quantum Field Theory with action (\ref{action})
could have been solved directly by using the equations for the boundary
$S$--matrix found by Ghoshal and Zamolodchikov \cite{goza}.
Indeed, once the
expansion of the field $\varphi$ for $x\geq0$
\begin{equation}
\varphi(x,t)=\frac{1}{\sqrt{2}}\int\frac{\mbox{d}\theta}{2\pi}
\left[A(\theta)e^{-im(t\cosh\theta-x\sinh\theta)}+
A^\dagger(\theta)e^{im(t\cosh\theta-x\sinh\theta)}\right]
\end{equation}
is inserted into the boundary condition
\begin{equation}
\partial_x\varphi|_{x=0}=c\,\varphi|_{x=0}\, ,
\end{equation}
we have the equation
\[
{\cal{B}}\,A^\dagger(-\theta)=R(\theta){\cal{B}}\,A^{\dagger}(\theta)\, ,
\]
with
\begin{equation}
R(\theta)=\frac{\sinh\theta-ic/m}{\sinh\theta+ic/m}
\end{equation}
and ${\cal{B}}$ the boundary operator.
The Reflection amplitude $R(\theta)$ can be used to define the boundary state
(\ref{bst}) and the Green's function (\ref{last}). However, by using
this approach, the different role played by the Transmission and Reflection
amplitudes of the quantum defect line would have been missed.

As our last remark, it is worth to mention that the action (\ref{action})
has been extensively studied in the context of phase transitions
near surfaces in \cite{BM,surf}, where it describes the {\it high--temperature}
Landau--Ginzburg lagrangian for a magnetic system with boundary. Its
validity is restricted by the occurrence of a surface phase transition for
$c<0$: high--temperature then means temperature higher than the surface
critical temperature, which is $\tau_s=0$ (i.e. $T_s=T_c$) if $c>0$ and
$\tau_s=|c|^2$ if $c$ is negative. Notice that these limits are those
mentioned
after equation (\ref{F}) and were also discussed in the context of boundary
(or defect) scattering amplitudes in \cite{dms}. In the last context,
the poles of $K(\theta)$ for $\tau \leq \tau_s$ simply imply spontaneous
emission
of pairs of particles from the boundary, a condition that destroys the
stability of the system.

\subsection*{Acknowledgments}

The author would like to thank G. Mussardo for the many useful discussions
and helps.

\pagestyle{empty}

\newpage

\begin{picture}(300,400)
\put(100,200){\begin{picture}(200,200)
\put(99,10){\line(0,1){200}}
\put(101,10){\line(0,1){200}}
\put(100,10){\line(0,1){200}}
\put(100,0){\makebox(0,0){$ {\cal{D}}$}}
\put(30,35){\vector(1,1){40}}
\put(70,75){\line(1,1){29}}
\put(70,53){\makebox(0,0){$A^{\dagger}(\beta)$}}
\put(99,106){\vector(-1,1){40}}
\put(59,146){\line(-1,1){29}}
\put(59,175){\makebox(0,0){$R(\beta)$}}
\put(101,106){\vector(1,1){40}}
\put(141,146){\line(1,1){29}}
\put(141,175){\makebox(0,0){$T(\beta)$}}
\end{picture}}
\end{picture}

\vspace{1mm}
\begin{center}
Figure 1 \\
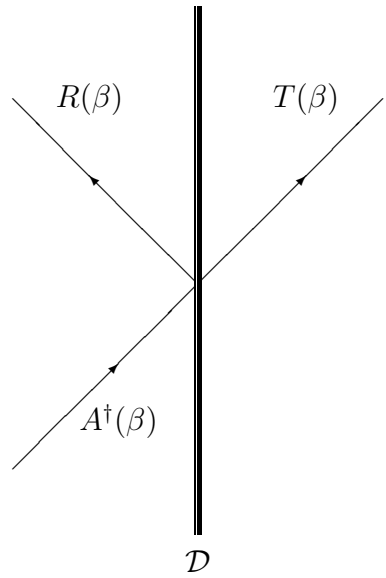
The particle hits the defect line and is scattered \\
according to the Reflection and Transmission amplitudes.
\end{center}

\newpage

\begin{picture}(500,300)
\put(30,0){\begin{picture}(400,300)
\put(50,149){\line(1,0){250}}
\put(50,150){\line(1,0){250}}
\put(50,151){\line(1,0){250}}
\multiput(175,151)(0,5){15}{\line(0,1){3}}
\multiput(175,149)(0,-5){15}{\line(0,1){3}}
\put(200,220){\makebox(0,0){$\beta$}}
\put(175,150){\vector(1,1){30}}
\put(205,180){\line(1,1){30}}
\put(115,90){\vector(1,1){30}}
\put(145,120){\line(1,1){30}}
\end{picture}}
\end{picture}
\vspace{1mm}
\begin{center}
\hspace{5mm}
Figure 2.a\\
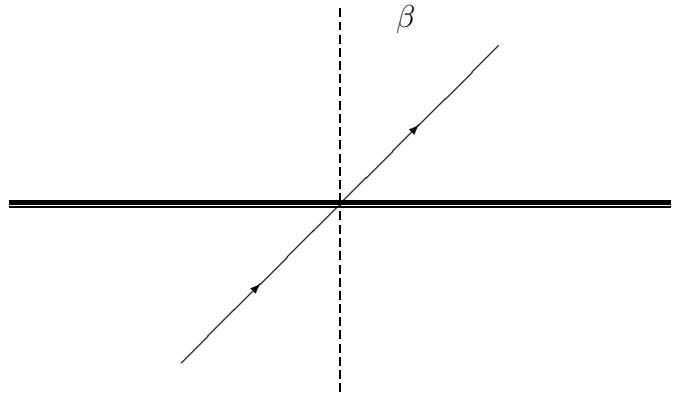
The process of transmission \\
with the defect line at $it=x=0$.
\end{center}

\newpage
\begin{picture}(500,300)
\put(30,0){\begin{picture}(400,300)
\put(50,149){\line(1,0){250}}
\put(50,150){\line(1,0){250}}
\put(50,151){\line(1,0){250}}
\multiput(130,152)(0,5){16}{\line(0,1){3}}
\put(165,220){\makebox(0,0){$\beta$}}
\put(95,220){\makebox(0,0){$-\beta$}}
\put(130,150){\vector(1,1){30}}
\put(130,150){\vector(-1,1){30}}
\put(160,180){\line(1,1){30}}
\put(100,180){\line(-1,1){30}}
\put(140,90){\vector(1,1){30}}
\put(170,120){\line(1,1){30}}
\put(260,90){\vector(-1,1){30}}
\put(230,120){\line(-1,1){30}}
\end{picture}}
\end{picture}
\vspace{1mm}
\begin{center}
\hspace{5mm}
Figure 2.b\\
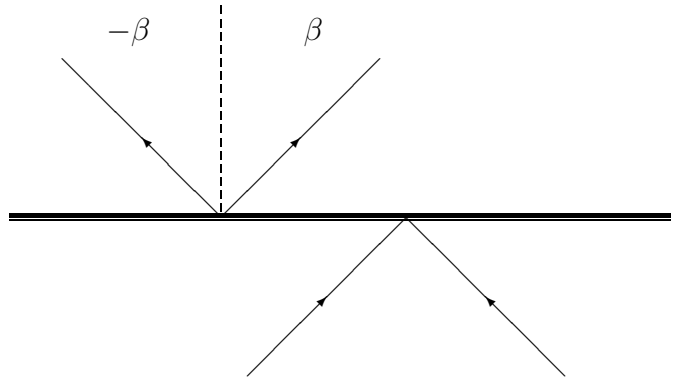
The process of reflection with \\
the defect line at $it=x=0$.
\end{center}

\end{document}